\documentclass[10pt,twocolumn,letterpaper]{article}

\usepackage{cvpr}              

\usepackage{graphicx}
\usepackage{amsmath}
\usepackage{amssymb}
\usepackage{booktabs}
\usepackage{siunitx}
\usepackage{lipsum}

\usepackage[pagebackref,breaklinks,colorlinks]{hyperref}

\usepackage[capitalize]{cleveref}
\crefname{section}{Sec.}{Secs.}
\Crefname{section}{Section}{Sections}
\Crefname{table}{Table}{Tables}
\crefname{table}{Tab.}{Tabs.}

\def\cvprPaperID{*****} 
\def\confName{CVPR}
\def\confYear{2023}

\begin{document}

\title{Screening Mammography Breast Cancer Detection}

\author{Debajyoti Chakraborty\\
Northeastern University\\
360 Huntington Ave, Boston, MA 02115\\
{\tt\small chakraborty.de@northeastern.edu}
}
\maketitle

\begin{abstract}
	Breast cancer is a leading cause of cancer-related deaths, but current programs are expensive and prone to false positives, leading to unnecessary follow-up and patient anxiety. This paper proposes a solution to automated breast cancer detection, to improve the efficiency and accuracy of screening programs. Different methodologies were tested against the \texttt{RSNA} dataset of radiographic breast images of roughly 20,000 female patients and yielded an average validation case pF1 score of 0.56 across methods.
\end{abstract}

\section{Introduction}
\label{sec:intro}

Breast cancer is one of the most commonly occurring cancer in the world with 2.3 million new breasts cancer diagnoses and 685,000 deaths in 2020 alone \cite{who-article-breast-cancer}. Although the mortality rate in developed nations have dropped by 40\% over the last 40 years due to regular mammography screening programs, such in not the case in many other countries due a looming shortage of radiologists.
\par
As with any cancer or disease in general, early detection and treatment is critical to reducing complications and fatalities. However, currently such procedures require the expertise of highly-trained human observers, primarily radiologists, making the overall process expensive to conduct and prone to human error, worsening the problem.

\section{Problem Statement}
\label{sec:problem}

A major problem in mammography screening is that it often leads to a high incidence of false positive results. This is usually followed by further screening tests, inconvenient follow-up, and sometimes, unneeded tissue sampling (needle biopsy) which may lead to further unrelated complications, causing unnecessary anxiety.
\par
This paper aims to improve the automatic detection of breast cancer in screening mammograms obtained from regular screening programs, with the goal being to reduce the occurrences of false positives in a clinical setting.

\subsection{Dataset}
\label{subsec:data}

\noindent\textbf{\footnotesize [This dataset contains radiographic breast images of female subjects.]}
\par
The dataset \cite{rsna-breast-cancer-detection} has been generously provided by the Radiological Society of North America (RSNA). RSNA is a non-profit organization that represents 31 radio-logic sub-specialties from 145 countries around the world.
\par
It contains radiographic breast images of roughly 20,000 female patients with usually four images per patient with two lateral \textbf{[left, right]} images per view \textbf{[mediolateral-oblique (MLO), crainal-caudal (CC)]}.
\begin{table}[b]
	\caption{Metadata for each patient and image}
	\centering
	\begin{tabular}{p{0.33\linewidth} p{0.58\linewidth}}
		\toprule
		\texttt{site id} & ID code for the source hospital \\
		\texttt{machine id} & ID code for the imaging device \\
		\texttt{patient id} & ID code for the patient \\
		\texttt{image id} & ID code for the respective image \\
		\texttt{laterality} & whether the image is of the left or right breast \\
		\texttt{view} & orientation of the image \\
		\texttt{age} & patient’s age in years \\
		\texttt{implant} & whether the patient had breast implants at the patient level \\
		\midrule
		\texttt{density} & rating for how dense the breast tissue is, with A being the least dense and D being the most dense \\
		\texttt{biopsy} & whether a follow-up biopsy was performed on the breast \\
		\texttt{invasive} & whether or not the cancer (if true) proved to be invasive \\
		\texttt{BIRADS} & 0 if the breast required follow-up, 1 if the breast was rated as negative for cancer, and 2 if the breast was rated as normal \\
		\texttt{difficult negative case} & true if the case was unusually difficult to diagnose \\
		\midrule
		\texttt{cancer} & whether or not the breast was positive for malignant cancer \\
		\bottomrule
	\end{tabular}
	\label{Tab:meta}
\end{table}
\begin{figure*}[h]
	\centering
	\includegraphics[width=\linewidth]{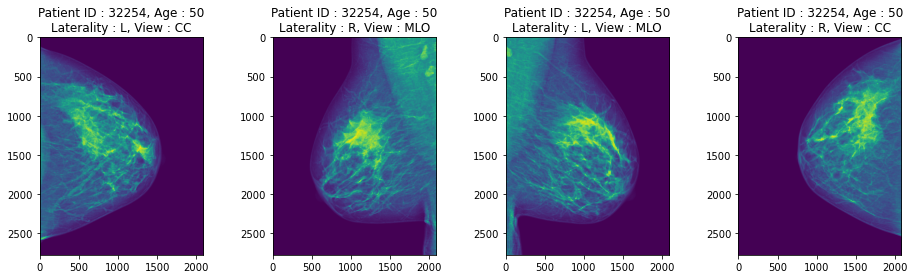}
	\caption{Example \texttt{MLO} and \texttt{CC} view of left(\texttt{L}) and right(\texttt{R}) breasts for patient id \texttt{32254}.}
	\label{fig:mlocc}
\end{figure*}
\par
The raw dataset contains around 54,700 mammography images in the Digital Imaging and Communications in Medicine (\texttt{DICOM}) \cite{dicom-innolitics-mammography} format. There is a significant class imbalance in the target variable \texttt{cancer}, with 1,158 data-points for the positive class and 53,548 data-points for the negative class. For the purpose of this paper, \texttt{image id} serves as the input data, \texttt{cancer} as the binary target, along with only \texttt{age} and \texttt{implant} information as additional metadata. The rest of the metadata will be included in a future implementation for better inference.

\section{Approach}
\label{sec:approach}

\subsection{Image encoding}
\label{subsec:image}
Majority of the work was involved in pre-processing the \texttt{DICOM} images and converting them into \texttt{png} files for easier post-processing and training. Most of the dicoms in the dataset contained \texttt{JPEG2000} encoded images. The bitstream was extracted and decoded on the GPU, saved as \texttt{png} files. Although this process was not lossless, it saved a lot of overhead for processing them individually, as they were quite high-dimensional.

\subsection{Pre-processing}
\label{subsec:preprocess}

Different pre-processing techniques were tried to improve the chances of accurate classification. Two of the primary techniques were Region of Interest \texttt{ROI} cropping and normalization. However, upon further research, it was only decided to use normalization and drop cropping, due to the variability in image dimensions and loss of information by resizing. Photometric interpretation, an attribute that specifies the intended interpretation of the raw pixel data were either \texttt{MONOCHROME1} or \texttt{MONOCHROME2}. \texttt{MONOCHROME1} is usually used when the mammogram is intended to be viewed in a \textbf{white} background and \texttt{MONOCHROME2} is usually used when the mammogram is intended to be viewed in a \textbf{black} background. The author chose to invert all \texttt{MONOCHROME1} images and keep all \texttt{MONOCHROME2} images intact, as they display more lesion-based information when viewed by a machine. All original image arrays were normalized to between 0 and 1 for uniformity, and resized into $512 \times 512$ gray-scale images for consistency.
\begin{figure}[h]
	\centering
	\includegraphics[width=\linewidth]{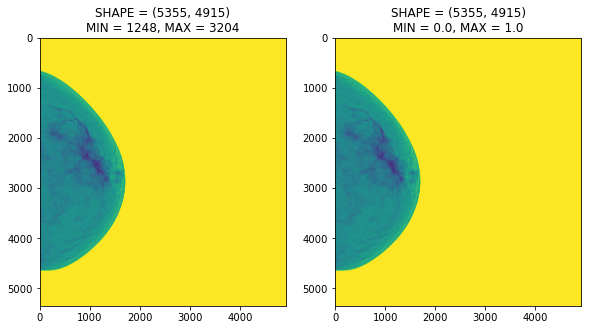}
	\caption{(left) Random sample of a \texttt{MONOCHROME1} image showing the minimum pixel value and (right) the same image after inverting and normalizing all pixel values to between 0 and 1.}
	\label{fig:dicom}
\end{figure}

\subsection{Feature extraction}
\label{subsec:feature}

An \texttt{EfficientNetV2} \cite{tan2021efficientnetv2} with pre-trained weights was used to extract essential features from the images, encoding them to a $1000$-dimensional feature vector. It was developed by researchers at Google \cite{mingxing-tan-2019-google-ai} that achieved improved accuracy and efficiency than conventional Convolutional Neural Networks (CNNs), lowering parameter size and execution time by an order of magnitude.
\par
An modification of the dataset was later prepared by combining the exported image feature vectors with two meta features: normalized \textbf{age} and \textbf{implant} information. For the purpose of improving predictions, a combination of synthetic under-sampling and over-sampling techniques were also tested for performance. \cite{Buda_2018}.

\subsection{Classification methods}
\label{subsec:method}

These popular machine learning techniques for classification were trained on this data to derive inference.

\subsubsection{Logistic Regression (LR)}
\label{logr}

A simple LR model with \texttt{L2} regularization was trained on the unbalanced dataset with individually assigned class weights and has been described below.
\par
Given $X_i$ as the input feature vector and $y_i \in \{0,1\}$ as the target variable for data point $i$, the probability of the positive class $y_i=1$ was predicted by the fitted model as:
\begin{equation}
	\hat{p}(X_i) = P(y_i=1|x) = \frac{1}{1 + \exp(-X_i w - w_0)}
\end{equation}
The cost function that was minimized using a solver was:
\begin{multline}
	\min_w C \sum_{i=1}^{Nn} (-y_i log(\hat{p}(X_i)) - \\
	(1-y_i)\log(1-\hat{p}(X_i))) + \frac{1}{2} \|x\|_2^2
\end{multline}

\subsubsection{Support Vector Machine (SVM)}
\label{svm}

For implementing a SVM model, a scalable, input data independent polynomial kernel approximation method \cite{10.1145/2487575.2487591} was used. Given $x$, $y$ as input features and $d$ as the polynomial kernel degree, a simple kernel function used was:
\begin{equation}
	k(x, z) = (\gamma x^\top z + c_0)^d
\end{equation}
Next, given that $X_i \in \mathbb{R}^d$ and $y_i \in \{-1,1\}$ as the target variable for data point $i$, the following problem was solved:
\begin{equation}
	\begin{cases}
		\min_{w,b,\xi} \frac{1}{2} \| w \| + C \sum_{i=1}^{n} \xi_i \\
		\text{such that } y_i (w^\top k(x, z) + b) \geq 1 - \xi_i \\
		\xi_i \geq 0, \quad \forall \quad i=1,\cdots,n
	\end{cases}
\end{equation}
Equivalently formulated, it was written as:
\begin{equation}
	\min_{w,b} \frac{1}{2} \| w \| + C \sum_{i=1}^{n} \max(0, 1 - y_i(w^\top k(x, z) + b))
\end{equation}

\subsubsection{Breast-level single-view-single-laterality model}
\label{dnn}

A simple deep-neural network was designed to perform inference on the ensemble of features described above.
\begin{itemize}
	\item \textbf{Input:} Each $1000 \times 1$ vector was fed into an \textbf{image encoder} part of the network and after two hidden dense layers was concatenated with the feature outputs from a \textbf{metadata encoder}.
	
	\item \textbf{Activation:} Each intermediate layer had a \textbf{ReLU} activation function, except for the output layer, where a \textbf{Sigmoid} activation was used.
	\begin{equation}
		\texttt{ReLU: } f(z) = \max(0, z) \in [0,z]
	\end{equation}
	\begin{equation}
		\texttt{Sigmoid: } f(z) = \frac{1}{1 + \exp(-z)} \in [0,1]
	\end{equation}
	\item \textbf{Loss:} The loss with respect to the target variable was calculated using \textbf{Binary Cross-entropy (BCE)} loss, once while using categorical class weights and another without.
	\begin{multline}
		-\frac{1}{N} \sum_{i=0}^{N} y_i \log(\hat{y_i}) + (1 - y_i)\log(1 - \hat{y_i})
	\end{multline}
	\item \textbf{Class imbalance:} In the latter case, a \textbf{Straified Batch Sampling (SBS)} methodology was used when randomly sampling from the dataset during training.
	\begin{equation}
		\text{SBS} = \frac{\text{Total sample size}}{\text{Dataset population}} \times \text{Class population}
	\end{equation}
	\item \textbf{Optimizer:} \texttt{Adam} \cite{kingma2017adam} was chosen as the optimizer for our minimization problem. Given $\eta=0.0003$ as initial learning rate, $g_t$ as gradient at time $t$ along $w_j$, $\nu_t$ as exponential average of gradients along $w_j$, $s_t$ as exponential average of squares of gradients along $w_j$ and $\beta_1$, $\beta_2$ as hyper-parameters:
	\begin{equation}
		\nu_t = \beta_1 \nu_{t-1} - (1 - \beta_1) g_t
	\end{equation}
	\begin{equation}
		s_t = \beta_2 s_{t-1} - (1 - \beta_s) g_t^2
	\end{equation}
	\begin{equation}
		\partial w_t = -\eta \frac{\nu_t}{\sqrt{s_t + \epsilon}} g_t
	\end{equation}
	\begin{equation}
		w_{t+1} = w_t + \partial w_t
	\end{equation}
	\item \textbf{Metrics:} A range of different metrics were tried to understand the maximum efficiency of the model. As for a problem with a class imbalance, \textbf{accuracy} was unreliable because of the major bias towards negative class. In this regard, performances of binary area under the receiver operating characteristic curve (AUROC), binary precision, binary recall, binary $F_1$ score were compared. A comprehensive comparison of these metrics have been provided in Figure 3.
	\par
	The model was finally evaluated using the \textbf{probabilistic $F_1$ score ($pF_1$)} \cite{yacouby-axman-2020-probabilistic} as this extension accepts probabilities instead of binary classifications. With $p_x$ as the probabilistic version of $X$:
	\begin{equation}
		pF_1 = 2\frac{P_\text{precision}P_\text{recall}}{P_\text{precision} + P_\text{recall}}
	\end{equation}
	\begin{equation}
		P_\text{precision} = \frac{P_\text{true positive}}{P_\text{true positive} + P_\text{false positive}}
	\end{equation}
	\begin{equation}
		P_\text{recall} = \frac{P_\text{true positive}}{\text{true positive} + \text{false negative}}
	\end{equation}
\end{itemize}
\begin{table}[h]
	\caption{Simple Single-view-single-laterality model architecture}
	\centering
	\begin{tabular}{l l l c}
		\toprule
		Layer & Input & Output & Parameters \\
		\midrule
		SimpleFCN & [,$1002$] & [,$1$] & -- \\
			\quad Sequential: 1-1 & [,$1000$] & [,$10$] & -- \\
				\qquad Linear: 2-1 & [,$1000$] & [,$100$] & $100,100$ \\
				\qquad ReLU: 2-2 & [,$100$] & [,$100$] & -- \\
				\qquad Linear: 2-3 & [,$100$] & [,$10$] & $1,010$ \\
				\qquad ReLU: 2-4 & [,$10$] & [,$10$] & -- \\
				\qquad Linear: 2-3 & [,$10$] & [,$1$] & $11$ \\
				\qquad Sigmoid: 2-4 & [,$1$] & [,$1$] & -- \\
			\quad Sequential: 1-2 & [,$2$] & [,$1$] & -- \\
				\qquad Linear: 2-5 & [,$2$] & [,$2$] & $6$ \\
				\qquad ReLU: 2-6 & [,$2$] & [,$2$] & -- \\
				\qquad Linear: 2-7 & [,$2$] & [,$1$] & $3$ \\
				\qquad Sigmoid: 2-8 & [,$1$] & [,$1$] & -- \\
			\quad Concatenate: 1-3 & [,$2$] & [,$1$] & -- \\
			\quad Sigmoid: 1-4 & [,$1$] & [,$1$] & -- \\
		\bottomrule
	\end{tabular}
	\label{Tab:model}
\end{table}
\par
The total parameters in the model was estimated at $101,130$, all of which were trainable. The total model size excluding the feature extractor was estimated at \qty{20.56}{\mega\byte}, with input size at \qty{16.42}{\mega\byte}, forward/backward pass size at \qty{3.74}{\mega\byte} and all parameter size at \qty{0.40}{\mega\byte}.
\begin{figure}[h]
	\centering
	\includegraphics[width=\linewidth]{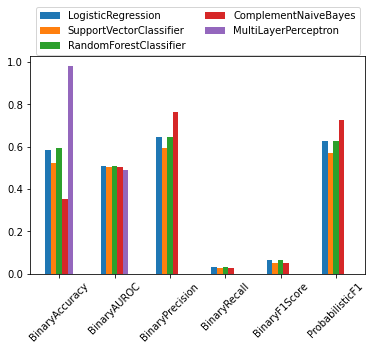}
	\caption{Comparison of different evaluation metrics.}
	\label{fig:eval}
\end{figure}

\section{Results}
\label{sec:result}

An extensive array of experiments were carried out to estimate the best machine learning model suited for this dataset. This not only included varying hyper-parameters and trying out different models, but also making efforts in transforming, augmenting and generating synthetic data points for the imbalanced classes.
\par

\begin{figure}[h]
	\centering
	\captionsetup{justification=centering}
	\includegraphics[width=\linewidth]{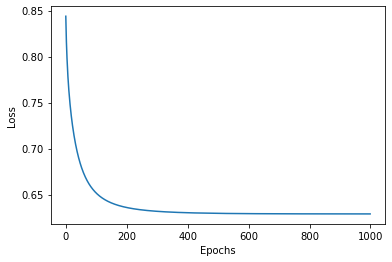}
	\caption{Validation loss graph trained for 1000 iterations with breast-level single-view-single-laterality DNN model.}
	\label{fig:eval2}
\end{figure}
\par
The results for $pF_1$ score were compared as well in order to have a more holistic view at all the approaches, and served well to gauge different model performances compared to each other and state of the art at present. This was due to lack to good test data, lack of sufficient positive classes, and personally, lack of time and resources.
\begin{table}[h]
	\caption{Probabilistic F-1 scores}
	\centering
	\begin{tabular}{l c}
		\toprule
		Model & $pF_1$ score \\
		\midrule
		State-of-the-art & $0.630$ \\
		Logistic Regression & $0.627$ \\
		Support Vector Machine & $0.572$ \\
		Random Forest Classifier & $0.626$ \\
		Complement Naive Bayes & $0.579$ \\
		Deep Neural Network & $0.481$ \\
		\bottomrule
	\end{tabular}
	\label{Tab:results}
\end{table}

\section{Summary}
\label{sec:summary}

As it can be inferred from the above table, the current state of any of the models is not better than $0.50$ probability, that is a $50\%$ chance of providing the correct class prediction, which is not any better than random chance. Comparing the results to the current state-of-the-art shows that the topic needs more correct predictions to be relevant.
\par
Other investigations that did not produce significant results include Region-of-interest (ROI) cropping, down-sampling, up-sampling or synthetically generating new samples for the dataset, and almost certainly leading to over-fitting. The rest of the metadata was not part of future test samples, so they were not included as well.
\par
Literature survey shows that the use of intense data augmentation pipelines and training models externally on similar datasets work well. This is mainly done to lessen the possibility of detection of false positives due to exposure to more data points in the positive class. There is also strong evidence that networks that capture both spatial and temporal information for a single patient, in a multi-view-multi-lateral model show drastically improved performance. The recent use of transformers for vision tasks has shown significant promise as well, but that calls for further investigation and is also outside the scope of this paper.

{\small
\bibliographystyle{ieee_fullname}
\bibliography{main}
}

\end{document}